\documentclass[conference, anonymous]{IEEEtran}
\IEEEoverridecommandlockouts
\usepackage{cite}
\usepackage{amsmath,amssymb,amsfonts}
\usepackage{algorithmicx}
\usepackage{graphicx}
\usepackage{textcomp}
\usepackage{comment}
\usepackage{xcolor}
\usepackage{multirow}
\usepackage{algorithm}
\usepackage{algpseudocode}
\usepackage{soul}
\usepackage[table]{xcolor}
\definecolor{rowgray}{gray}{0.92}  

\def\BibTeX{{\rm B\kern-.05em{\sc i\kern-.025em b}\kern-.08em
    T\kern-.1667em\lower.7ex\hbox{E}\kern-.125emX}}

\begin{document}

\title{RAG-targeted Adversarial Attack on LLM-based Threat Detection and Mitigation Framework}

\author{\IEEEauthorblockN{Seif Ikbarieh}
\IEEEauthorblockA{\textit{Department of Computer Science} \\
\textit{Tennessee Tech University}\\
Cookeville, USA \\
smikbarieh42@tntech.edu}
\and
\IEEEauthorblockN{Kshitiz Aryal}
\IEEEauthorblockA{\textit{School of Interdisciplinary Informatics} \\
\textit{University of Nebraska Omaha} \\
\textit{Omaha, USA} \\
karyal@nebraska.edu}
\and
\IEEEauthorblockN{Maanak Gupta}
\IEEEauthorblockA{\textit{Department of Computer Science} \\
\textit{Tennessee Tech University}\\
Cookeville, USA \\
mgupta@tntech.edu}
}
\maketitle

\begin{abstract}
The rapid expansion of the Internet of Things (IoT) is reshaping communication and operational practices across industries, but it also broadens the attack surface and increases susceptibility to security breaches. Artificial Intelligence has become a valuable solution in securing IoT networks, with Large Language Models (LLMs) enabling automated attack behavior analysis and mitigation suggestion in Network Intrusion Detection Systems (NIDS). Despite advancements, the use of LLMs in such systems further expands the attack surface, putting entire networks at risk by introducing vulnerabilities such as prompt injection and data poisoning. In this work, we attack an LLM-based IoT attack analysis and mitigation framework to test its adversarial robustness. We construct an attack description dataset and use it in a targeted data poisoning attack that applies word-level, meaning-preserving perturbations to corrupt the Retrieval-Augmented Generation (RAG) knowledge base of the framework. We then compare pre-attack and post-attack mitigation responses from the target model, \textit{ChatGPT-5 Thinking}, to measure the impact of the attack on model performance, using an established evaluation rubric designed for human experts and judge LLMs. Our results show that small perturbations degrade LLM performance by weakening the linkage between observed network traffic features and attack behavior, and by reducing the specificity and practicality of recommended mitigations for resource-constrained devices.
\end{abstract}


\section{Introduction}
The Internet of Things (IoT) represents a rapidly expanding ecosystem of interconnected devices that communicate across networks to enable data-driven automation and control. By the end of 2024, the number of active IoT devices utilized globally has reached 18.8 billion, marking a 13\% increase from 2023 \cite{iotanalytics2024}. This rapid proliferation highlights not only the pace of adoption but also the growing economic significance of IoT, with global revenues estimated at nearly \$1 trillion in 2024 and projected to surpass \$2 trillion annually by 2030 \cite{gsmaintelligence2025}. Within this broader landscape, the Industrial IoT (IIoT) integrates these connected systems into manufacturing, healthcare, transportation, and energy sectors, enabling predictive maintenance, real-time monitoring, and operational efficiencies critical to digital transformation.
Nevertheless, the rapid expansion of IoT applications introduces significant security concerns, as it widens the attack surface available to malicious actors. IoT devices are inherently vulnerable due to constrained hardware, outdated or unpatched firmware, weak authentication mechanisms, and default configurations that are rarely changed. According to SonicWall, IoT-targeted cyberattacks increased by 124\% in 2024 alone, underscoring the need to build robust systems to secure such environments \cite{sonicwall2025}.

The integration of Artificial Intelligence (AI) into cybersecurity has evolved from an emerging trend to a strategic necessity, strengthening defensive capabilities while reducing operational costs. According to IBM, organizations that deployed AI-driven security systems reduced the average breach cost by approximately \$1.9 million compared to those without such technologies, while also shortening the breach life-cycle by nearly 80 days \cite{ibm2025}. Recently, existing work has incorporated Large Language Models (LLMs) into Network Intrusion Detection Systems (NIDS), providing these systems with the ability to perform automated attack analysis and generate mitigation suggestions \cite{ikbarieh2025, 10.1145/3663408.3663424, juttner2023chatids}. 

However, integrating LLMs into NIDS broadens the attack surface, with risks like Retrieval-Augmented Generation (RAG) poisoning that corrupts knowledge bases and misleads retrieval \cite{zou2025poisonedrag,10580402}, as well as threats such as prompt injection, system prompt disclosure and jailbreak-style policy circumvention, and exploitation of hallucinated content \cite{owasp2025}. 
Despite these risks, existing LLM-based frameworks are rarely tested against adversarial attacks, especially under the resource constraints of IoT and IIoT deployments. The presence of entry points such as RAG data-poisoning and prompt injection highlights a critical \textbf{research gap} that remains underexplored in current frameworks, leaving open questions about reliability under black-box transfer attacks and retrieval corruption. Motivated by this gap, the contributions of this work are as follows:
\begin{itemize}
    \item We construct an IoT attack description dataset covering 18 different attack descriptions by leveraging prompt engineering.
    \item We test the adversarial robustness of a framework by carrying out a transfer-learning-based data-poisoning attack on its RAG component.
    \item We fine-tune a Bidirectional Encoder Representations from Transformers (BERT) model \cite{devlin2019bert} as a surrogate target using IoT attack descriptions. Then, using Textfooler \cite{jin2020bert}, we create word-level perturbations that are both semantic-preserving and constrained by Part-of-Speech (POS) tags. Finally, we inject these adversarially modified descriptions into the RAG knowledge base, which disrupts the retrieval process for downstream prompts sent to \textit{ChatGPT-5 Thinking} \cite{openai2025-thinking}.
    \item We evaluate the effectiveness of the adversarial attack by comparing the pre-attack and post-attack responses from our target \textit{ChatGPT-5 Thinking} model using a set of performance metrics. These responses are assessed both by human experts and by judge Large Language Models (LLMs) based on the specified metrics.
    \item Results show that \textit{ChatGPT-5 Thinking}'s performance is degraded after the attack, demonstrating the vulnerability of such LLM-based system towards RAG-poisoning attack
    
\end{itemize}
The remaining sections of this paper are structured as follows. Section \ref{sec:rel} reviews related work on LLM-based attack analysis and mitigation, as well as adversarial attacks on LLMs. Section \ref{sec:datasets} describes the datasets used in the framework. Section \ref{sec:frame} details the framework, including attack detection, RAG, prompt engineering, and the adversarial attack. Section \ref{sec:eval} presents the evaluation metrics and results. Finally, Section \ref{sec:summary} concludes the paper and discusses future research directions.

\section{Related works}
\label{sec:rel}

LLM-driven cyber defense has progressed from rule-based NIDS add-ons to end-to-end pipelines that pair classical detection with LLM-driven analysis and mitigation. Prior work integrates LLMs to explain alerts, extract indicators, and recommend mitigations. In \cite{ikbarieh2025}, an LLM-based framework combines a Random Forest (RF) classifier for multi-class attack detection on Edge-IIoTset \cite{9751703} and CICIoT2023 \cite{s23135941} with RAG-powered prompts to produce device-aware mitigations, and introduces a quantitative scoring rubric for LLM performance evaluation. The authors in \cite{10.1145/3663408.3663424} present ShieldGPT, a system that employs LLMs to analyze the behavior of Distributed Denial of Service (DDoS) attacks and propose mitigation suggestions to counter them. ShieldGPT has four key components: attack detection, traffic representation, domain knowledge injection, and role representation, which enable LLM-driven network traffic analysis and mitigation suggestion. Based on GPT-4 as the backbone LLM, ShieldGPT uses prompt engineering methods to create explanatory analyses and deployable defense mechanisms designed for particular devices. The authors in \cite{juttner2023chatids} introduce ChatIDS, a framework that leverages LLMs to convert IDS alerts into natural language explanations and recommended security measures. Their system is designed for IoT environments such as smart homes and aims to make IDS outputs more accessible to non-expert users. The pipeline retrieves alerts generated by Suricata, Snort, and Zeek \cite{suricata2025, snort2025, zeek2025}, incorporates metadata in them, and later provides them to ChatGPT for analysis and mitigation suggestion.

Despite the effectiveness of these systems, prior work has shown that the use of LLMs in such pipelines expands the overall attack surface, particularly demonstrating their vulnerability to word-level perturbations \cite{zou2025poisonedrag, 10580402}. The authors in \cite{zou2025poisonedrag} introduce PoisonedRAG, formalizing knowledge-base poisoning for RAG by crafting a small number of malicious passages that satisfy a retrieval condition and a generation condition so the poisoned text is both retrieved for a target query and steers the model to a chosen answer. They report a 90\% attack success rate by inserting only five poisoned texts into a knowledge base containing millions of clean texts. The authors in \cite{10580402} propose LLM-ATTACK, a two-stage word-level method that ranks token importance and then replaces selected words with LLM-suggested synonyms under semantic and POS constraints to yield valid and natural adversarial examples. They evaluate on multiple datasets and report strong automatic attack success, with human and GPT-4 evaluations confirming high validity and low detectability of the generated texts.

These findings highlight the potential for small, meaning-preserving modifications and targeted knowledge poisoning to undermine retrieval-grounded reasoning. However, prior work does not examine word-level adversarial attacks in IoT and IIoT network settings, leaving the adversarial robustness of RAG-powered attack analysis and mitigation frameworks untested. Motivated by this gap, we test the adversarial robustness of the framework in \cite{ikbarieh2025} by proposing a transfer-learning based data-poisoning attack that aims to degrade the attack behavior analysis and mitigation suggestion performance of \textit{ChatGPT-5 Thinking} using word-level perturbations that corrupt its RAG component.

\begin{table*}[!t]
\caption{Attack Types in the Framework}
\label{tab:common_attacks}
\begin{center}
\begin{tabular}{|l|p{12.2cm}|}
\hline
\textbf{Attack Category} & \textbf{Attack Types} \\
\hline
\textbf{DDoS} & Transmission Control Protocol Synchronization Flood (TCP SYN Flood) [TCP Flood, SYN Flood], User Datagram Protocol Flood (UDP Flood), Hypertext Transfer Protocol Flood (HTTP Flood), Internet Control Message Protocol Flood (ICMP Flood) \\
\hline
\textbf{Reconnaissance} & Port Scanning, Vulnerability Scanning, Operating System (OS) Fingerprinting/OS Scanning \\
\hline
\textbf{Spoofing} & MITM [Address Resolution Protocol (ARP) Spoofing, Domain Name System (DNS) Spoofing] \\
\hline
\textbf{Injection} & Cross-Site Scripting (XSS), Structured Query Language (SQL) Injection, Uploading \\
\hline
\textbf{Malware} & Backdoor, Password Cracking/Dictionary Brute Force \\
\hline
\end{tabular}
\end{center}
\end{table*}

\section{Datasets}
\label{sec:datasets}

We use two popular edge IoT/IIoT datasets for the evaluation in our work.
The Edge-IIoTset dataset \cite{9751703} was collected from a multi-layer testbed with diverse IoT and IIoT devices, sensors, and communication protocols to emulate real-world networks. It includes both benign and malicious traffic across 14 attack types that are organized into five categories, including Distributed Denial of Service (DDoS), information gathering, Man-in-the-Middle (MITM), injection, and malware. From the original 1,176 features, 61 were retained to represent traffic behavior, system processes, and protocol interactions. These features consist of both categorical and numerical values, making the dataset suitable for a range of Machine Learning (ML), Deep Learning (DL), and LLM-based methods. The testbed incorporated consumer IoT devices, industrial controllers, and networking components such as Raspberry Pi nodes, Modbus controllers, and SDN routers, providing a realistic environment for IoT intrusion detection research.

The CICIoT2023 dataset \cite{s23135941} was generated from a large-scale smart home testbed containing 105 IoT devices, including speakers, cameras, sensors, and hubs. It consists of normal traffic as well as 33 attack types organized into seven categories, including DDoS, DoS, reconnaissance, web-based, brute force, spoofing, and Mirai. The dataset provides 47 features extracted from packet flows that capture traffic patterns and protocol-level attributes. These features also include both categorical and numerical values, which support diverse approaches ranging from classical ML to modern DL architectures. Devices such as Google Nest Mini, Philips Hue Bridge, and D-Link cameras contribute to its realism, making it a valuable resource for evaluating NIDS in IoT networks. 

The framework uses the Edge-IIoTset and CICIoT2023 datasets for attack detection, LLM prompting, and benchmarking of LLM responses. Both datasets are aligned on 13 common attack types, distributed across a total of 28 classes, as shown in Table \ref{tab:common_attacks}, with square brackets indicating one-to-many mappings. This alignment harmonizes synonymous labels across datasets and preserves finer variants where necessary, enabling consistent prompting and fair cross-dataset comparisons.

\section{RAG-targeted Adversarial Attack on LLM-based Framework}
\label{sec:frame}
\begin{figure*}[!t]
    \centering
    \includegraphics[width=1\linewidth]{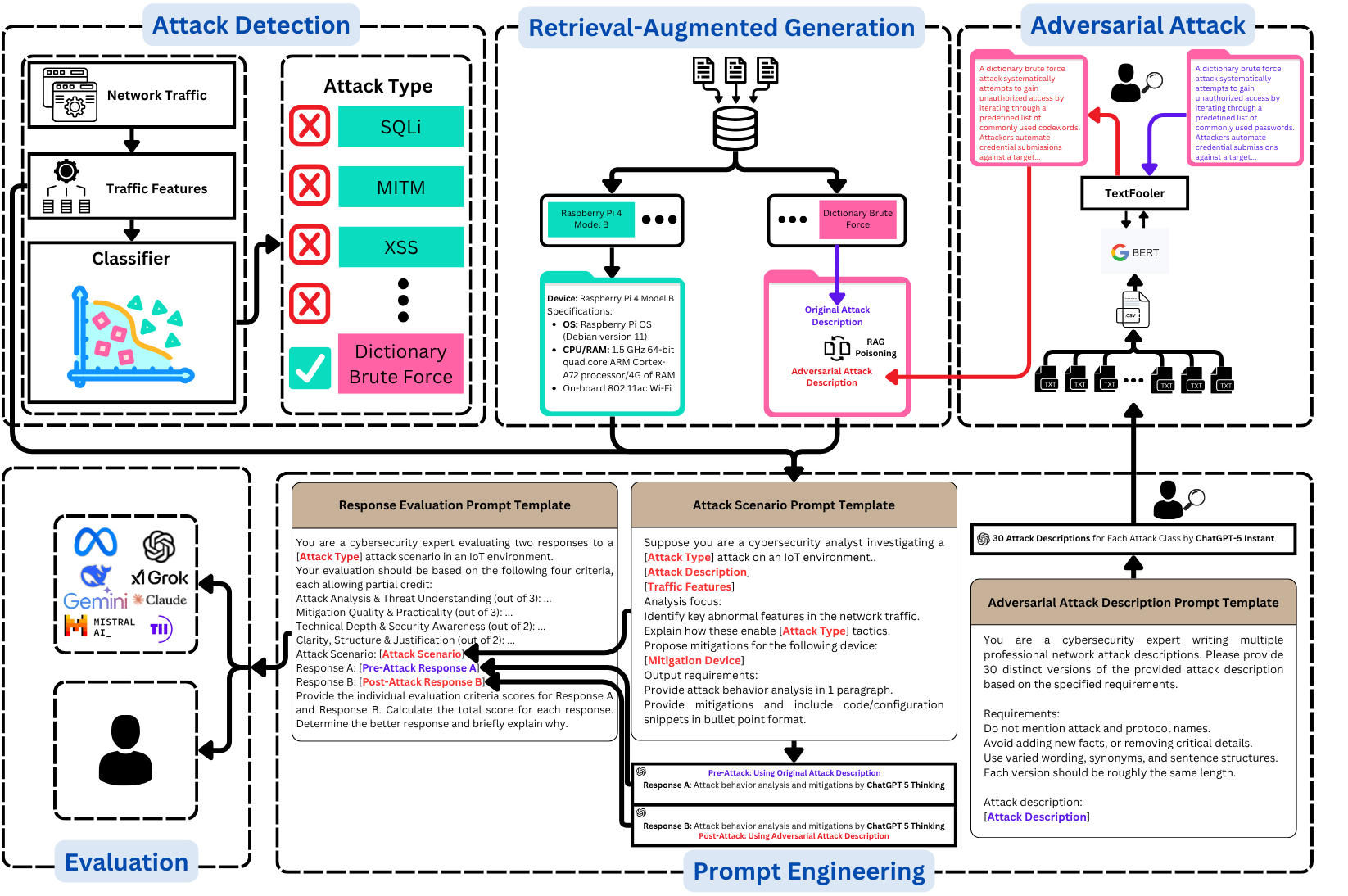}
    \caption{RAG-targeted Adversarial Attack on LLM-based Framework Overview}
    \label{fig:Framework-Diagram}
\end{figure*}

Figure \ref{fig:Framework-Diagram}, illustrates the framework design, which integrates an RF classifier for attack detection with \textit{ChatGPT-5 Thinking} for attack analysis and mitigation suggestion, and highlights where our adversarial attack is integrated within the LLM-based component. The framework consists of five main components. The \textit{attack detection} component classifies network traffic into benign or malicious classes using a trained RF classifier. The \textit{RAG} component provides attack descriptions and device information to enrich the context of LLM prompts. The \textit{adversarial attack} component uses \textit{TextFooler} to generate adversarially perturbed attack descriptions against \textit{BERT}, a surrogate target model, and uses them to test the robustness of \textit{ChatGPT-5 Thinking}, the target model, utilizing the transferability of the attack. The \textit{prompt engineering} component organizes and delivers prompts to the LLMs for attack analysis, mitigation suggestion, evaluation by judge models, and the fine-tuning of the adversarial attack surrogate target model. Lastly, the \textit{evaluation} component assesses the quality of LLM-generated responses before and after the adversarial attacks on RAG, using predefined scoring metrics applied by both human experts and judge LLMs. Our framework is implemented in Python, and all experiments were conducted on an Ubuntu 22.04 server running on an Intel(R) Xeon(R) Gold 5320 CPU @ 2.20GHz and an NVIDIA A100 40GB GPU.

\subsection{Attack Detection}
The attack detection component begins by capturing raw network traffic and preprocessing it into structured features that describe protocol behavior, packet statistics, and flow characteristics. Data preprocessing includes removing irrelevant columns, handling missing values, eliminating duplicates, encoding categorical variables, \textit{standardizing} numerical features, and applying feature scaling to ensure consistent and high-quality inputs. These processed features are then fed into the RF classifier, which has been trained to distinguish between benign traffic and all attack classes across both datasets. 

The RF model was chosen for this component because it achieved the highest accuracy and F1 score on both the Edge-IIoTset and CICIoT2023 datasets. RF also handles heterogeneous feature types effectively, and remains computationally efficient for deployment in resource-constrained environments. This design allows the attack detection component to provide reliable multi-class attack detection while maintaining practicality in real-world IoT settings.

\subsection{Retrieval-Augmented Generation}
The RAG component retrieves context-specific information from an external knowledge base, providing the detected attack's description and mitigation device specifications. In the knowledge base, each attack label is mapped to a concise technical description of its behavior and impact, while device entries contain hardware and software details such as CPU, memory, operating system, and network interfaces. RAG anchors the LLM to the correct attack description and the target device context, providing a basis for accurate attack behavior analysis and context-aware mitigation suggestion. By retrieving only evidence aligned with the detected threat and the deployment environment, it reduces model hallucinations \cite{fayyazi2024proverag}, which preserves accurate reasoning. This grounding also supports defensible, device-appropriate controls and yields outputs that are more consistent and reproducible across runs.

All entries are embedded into dense vector representations using the all-MiniLM-L6-v2 sentence transformer \cite{wang2020minilm} and indexed with Facebook AI Similarity Search (FAISS) \cite{facebook2017faiss} for efficient similarity search. Once an attack class is identified, the corresponding attack description and device specifications are retrieved and incorporated into the LLM prompt, ensuring that generated responses are aligned with both the detected threat and the target device’s context.

\subsection{Adversarial Attack}
\label{sec:AdversarialAttack}
\begin{figure*}[!t]
    \centering
    \includegraphics[width=1\linewidth]{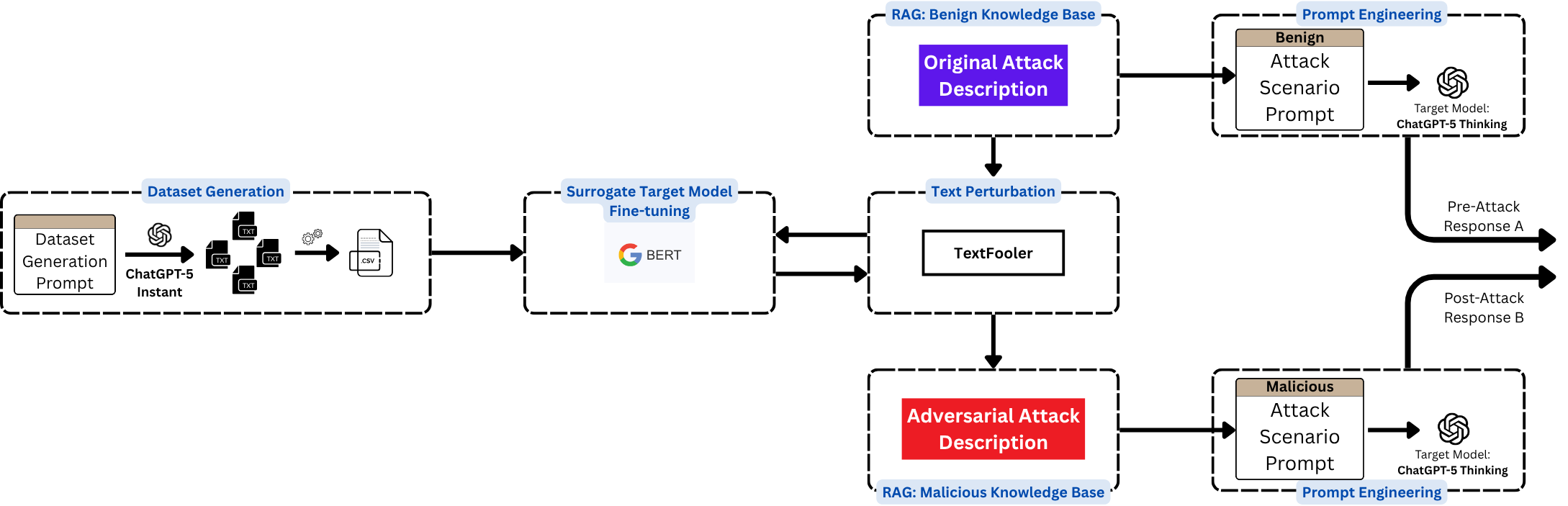}
    \caption{Adversarial Attack Pipeline Overview}
    \label{fig:adv_attack_overview}
\end{figure*}

Adversarial attack component, shown in Figure \ref{fig:adv_attack_overview} and summarized in Algorithm \ref{alg:adv_attack_alg1}, implements a transfer-learning-based data-poisoning attack to target the RAG knowledge base used by the attack scenario prompt. Transfer-learning is necessary because the target, \textit{ChatGPT-5 Thinking}, is a black-box model, and therefore fine-tuning the \textit{BERT}, a surrogate target model on paraphrased descriptions approximates its decision boundary and yields perturbations that transfer to the retrieval stage \cite{demontis2019adversarial}. The goal of this attack is to degrade the performance of attack behavior analysis and mitigation suggestion from \textit{ChatGPT-5 Thinking}, the target model, by clouding its retrieval context using adversarial attack descriptions that correspond to an incorrect attack class. As demonstrated in Figure \ref{fig:adv_attack_overview}, first, the attack description dataset is constructed by aggregating the paraphrased variants generated by \textit{ChatGPT-5 Instant} as discussed in Section \ref{sec:AttackDescriptionDatasetGeneration} into a Comma-Separated Values (CSV) file, with each description labeled by its corresponding attack class. Next, the combined samples are shuffled and split 80/20 into training and test sets and used to fine-tune a \textit{BERT} classifier for text classification. Fine-tuning a \textit{BERT} classifier on the paraphrased corpus produces a model that encodes how different wordings of the same attack map to label predictions. In other words, its decision boundary captures the text-to-label relationships present in the dataset. By crafting perturbations that change the surrogate model's prediction, we can identify minimal, meaning-preserving edits that exploit those decision boundaries.

The fine-tuned \textit{BERT} is then used as the surrogate target model for a word-level attack based on the \textit{TextFooler} algorithm \cite{jin2020bert}. \textit{TextFooler} identifies the tokens that most strongly influence \textit{BERT}’s classification decisions and iteratively replaces them with semantically similar alternatives while enforcing constraints from the Universal Sentence Encoder (USE) to preserve meaning and Part-of-Speech (POS) tagging to maintain grammaticality. The algorithm terminates once the perturbed text causes the fine-tuned \textit{BERT} model to misclassify the attack class. In this work, we use TextAttack's implementation of the TextFooler algorithm \cite{morris2020textattack}.

Prior to perturbation, each candidate description is verified to be correctly classified by the surrogate target model. We apply \textit{TextFooler} to the original attack descriptions in the framework, producing adversarially perturbed variants whenever \textit{BERT} misclassifies them. These adversarial descriptions then replace the corresponding entries in the RAG attack description knowledge base. During inference, when the attack scenario prompt queries the RAG component, the poisoned entries are retrieved instead of the originals, enabling the framework to evaluate their effect on \textit{ChatGPT-5 Thinking}’s attack analysis and mitigation suggestion.

\subsection{Prompt Engineering}
After an attack is detected by the RF classifier in \textit{attack detection} phase, the next stage leverages LLMs through prompt engineering to generate detailed analyses, mitigation suggestions, and performance evaluations using role-play prompts. In addition, we introduce a new prompt that is designed to create an attack description dataset for use in adversarial testing.

\subsubsection{Attack Description Dataset Generation}
\label{sec:AttackDescriptionDatasetGeneration}
Our \textit{dataset generation prompt} is designed for \textit{ChatGPT-5 Instant}, the successor to \textit{ChatGPT-4o} \cite{openai2025-thinking}. This model was chosen as the generator due to its demonstrated strength in cybersecurity knowledge, particularly IoT network security, as reflected in the CyberMetric benchmark \cite{10679494}. The goal of this prompt is to produce multiple paraphrased attack descriptions that preserve the original technical meaning of the original description while introducing controlled lexical and syntactic variation.

\begin{algorithm}[!t]
\caption{Adversarial Attack Pipeline}
\textbf{Input:} Set of original attack descriptions $\{d_c\}_{c \in C}$ with corresponding classes $C$\\
\textbf{Output:} Set of adversarial attack descriptions $\{d_c^{\mathrm{adv}}\}_{c \in C}$
\begin{algorithmic}[1]
\For{each class $c \in C$}
    \State Generate paraphrase variants $V_c$ for $d_c$ using ChatGPT\textendash5 Instant
\EndFor
\State Construct dataset $S \gets \bigcup_{c \in C} \{(v, c) \mid v \in V_c\}$
\State Fine-tune \textit{BERT} classifier $M$ on $S$ for multi-class text classification
\For{each class $c \in C$}
    \State $d_c^{\mathrm{adv}} \gets \text{TextFooler}(d_c, M)$
\EndFor
\For{each class $c \in C$ in the RAG attack description knowledge base}
    \State $d_c \gets d_c^{\mathrm{adv}}$
\EndFor\\
\Return None
\end{algorithmic}
\label{alg:adv_attack_alg1}
\end{algorithm}

The prompt begins by instructing the \textit{ChatGPT-5 Instant} to act as a cybersecurity expert that is tasked to produce 30 distinct versions of a given attack description while following a given set of requirements. Specifically, the generated descriptions must avoid explicitly naming attacks or protocols, preserve all critical technical details, and remain approximately the same length. This ensures consistency across variants while reducing redundancy. By enforcing varied word choices, synonyms, and sentence structures, the resulting dataset captures linguistic diversity without altering meaning.

\begin{figure}[!t]
\centering
\includegraphics[width=1\linewidth]{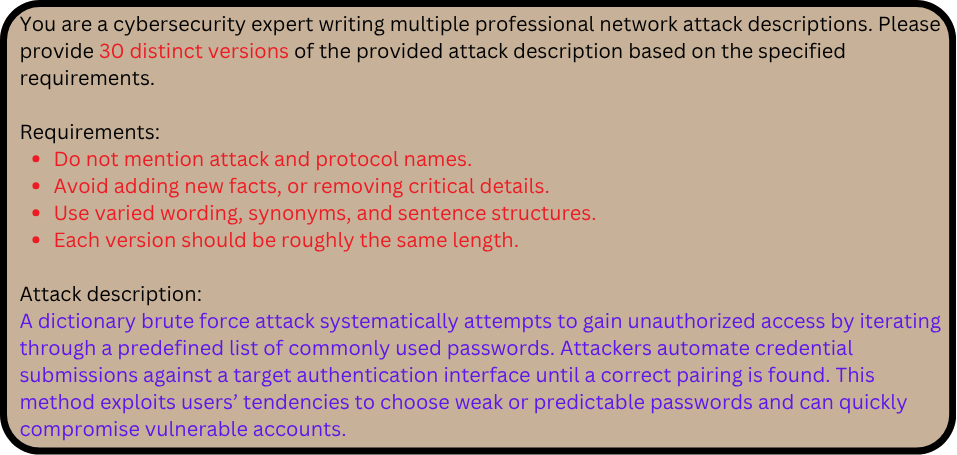}
\caption{Attack Description Dataset Generation Prompt}
\label{fig:Dataset-Generation-Prompt}
\end{figure}

\begin{figure}[!t]
\centering
\includegraphics[width=1\linewidth]{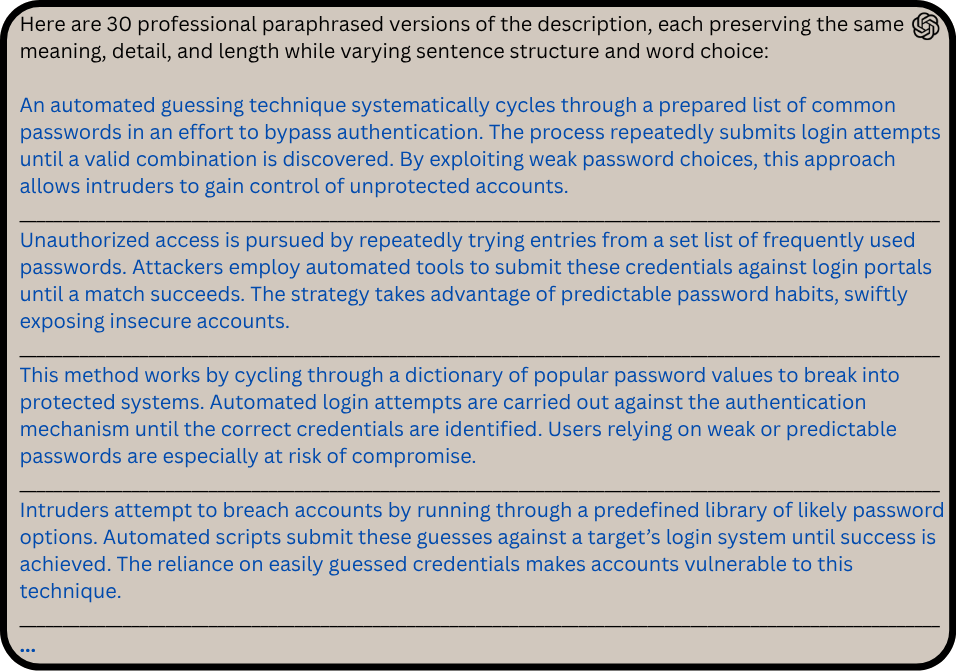}
\caption{Sample Attack Descriptions by \textit{ChatGPT-5 Instant}}
\label{fig:Dataset-Generation-Output}
\end{figure}

Figures \ref{fig:Dataset-Generation-Prompt} and \ref{fig:Dataset-Generation-Output} present the dataset generation prompt for a dictionary brute force attack and sample paraphrased outputs generated by \textit{ChatGPT-5 Instant}, respectively. We expand each attack type description from both datasets into multiple paraphrased variants, verify that the variants meet the requirements specified in the prompt, and save them into text files for use in the adversarial attack component.

\subsubsection{Attack Behavior Analysis and Mitigation}
\begin{figure}
    \centering
    \includegraphics[width=1\linewidth]{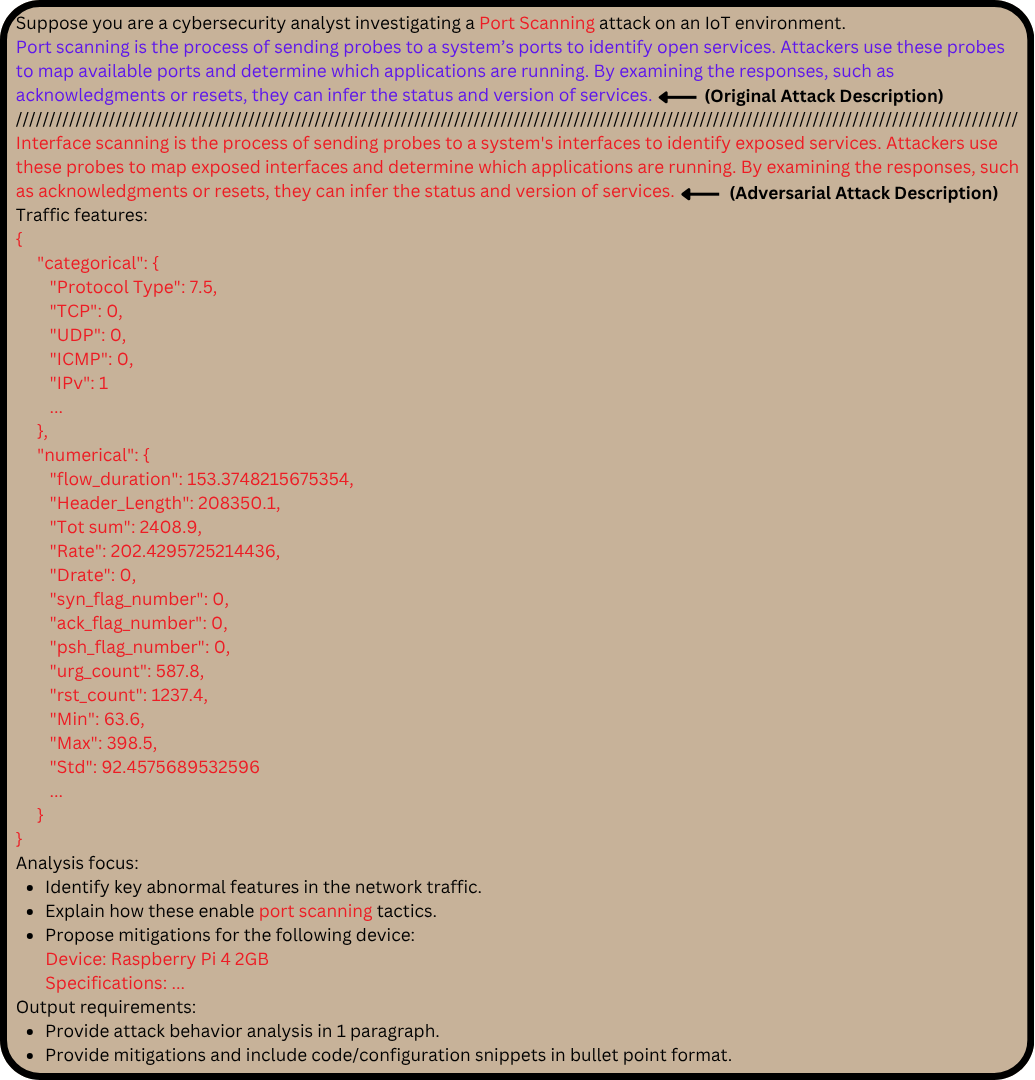}
    \caption{Example Attack Scenario Prompt}
    \label{fig:CIC-Example-Prompt}
\end{figure}

The \textit{attack scenario prompt} assigns the role of a cybersecurity analyst to a model that investigates a specific attack type in an IoT environment. The prompt includes the attack label classified by the RF model, JSON-formatted network traffic features from the attack snapshot, RAG-retrieved attack description and device specifications, as well as structured response requirements. An example in Fig. \ref{fig:CIC-Example-Prompt} shows a \textit{Port Scanning attack} from the CICIoT2023 dataset: the original description, its adversarially perturbed form (which the surrogate model in Section \ref{sec:AdversarialAttack} labeled as a \textit{Vulnerability Scanning attack description}), and the JSON-formatted traffic features and RAG-retrieved content. To avoid duplicating the full prompt, we present it once, noting that the original attack description is used for the pre-attack scenario, and the adversarial attack description is used for the post-attack scenario.

\begin{figure}
    \centering
    \includegraphics[width=1\linewidth]{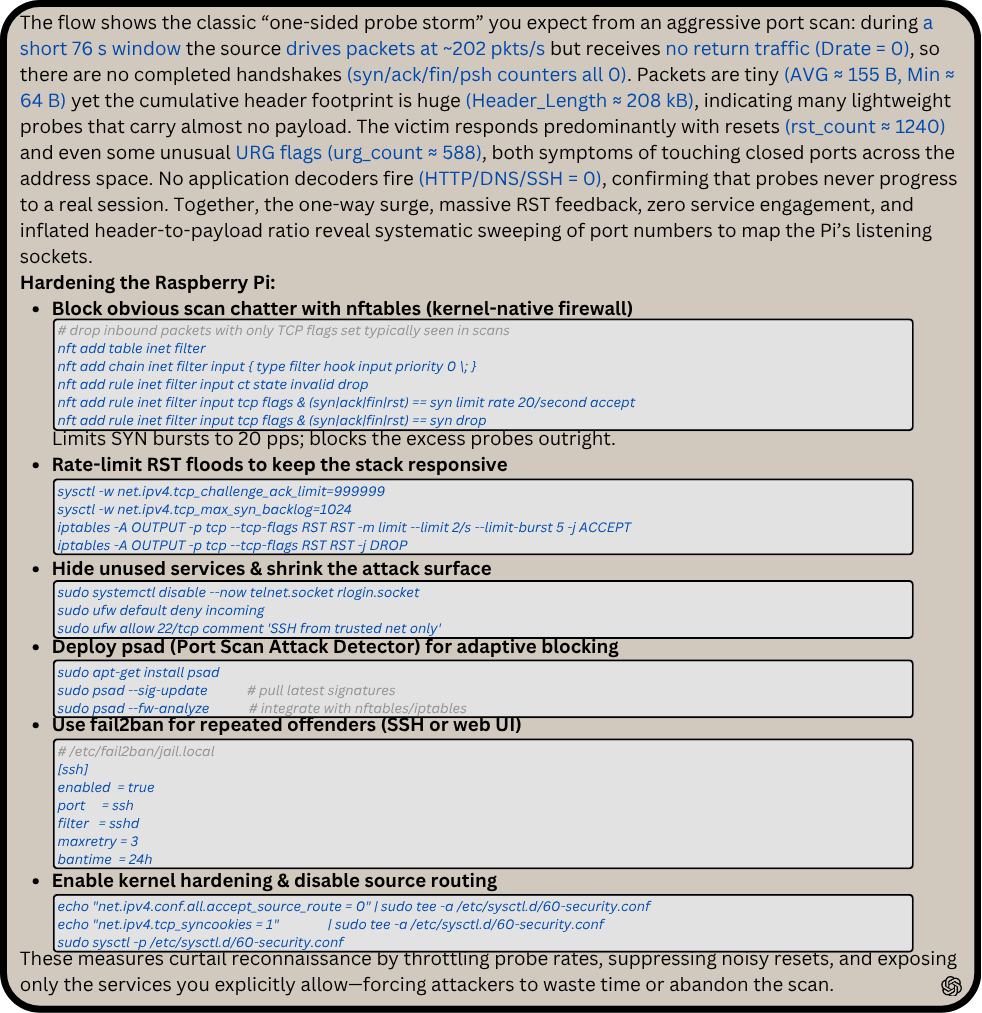}
    \caption{Pre-Attack Response by \textit{ChatGPT-5 Thinking}}
    \label{fig:CIC-Example-Output-Benign}
\end{figure}

\begin{figure}
    \centering
    \includegraphics[width=1\linewidth]{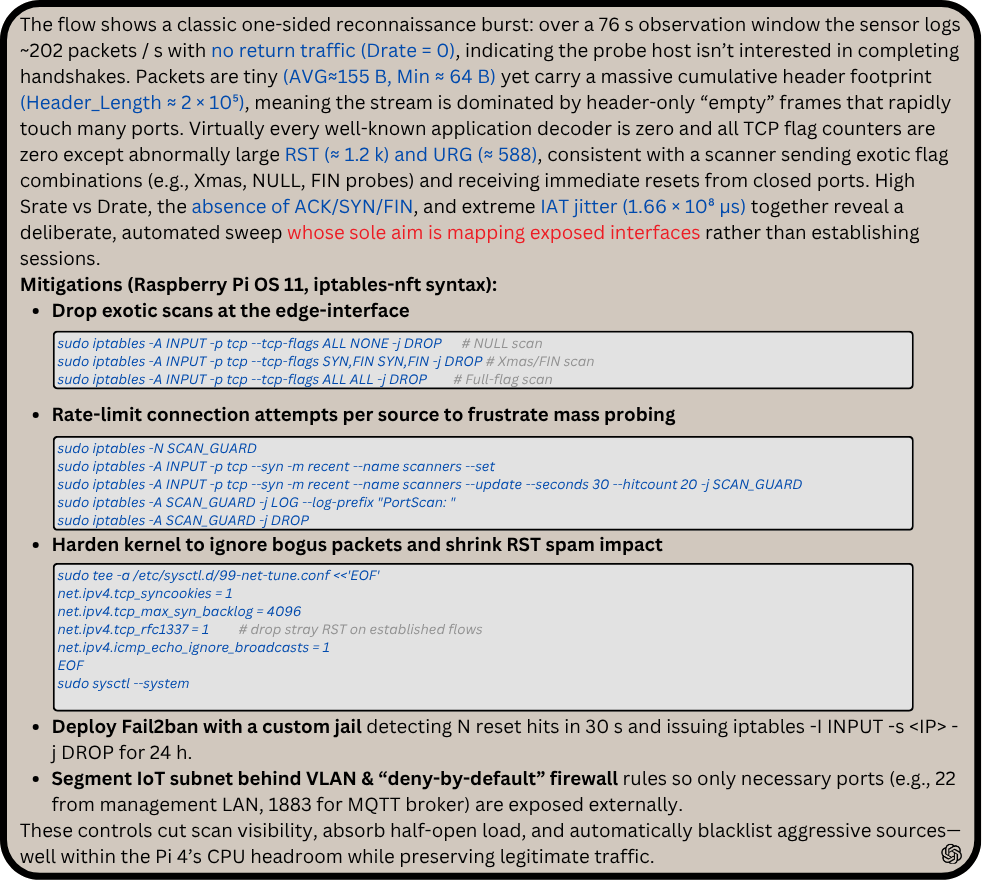}
    \caption{Post-Attack Response by \textit{ChatGPT-5 Thinking}}
    \label{fig:CIC-Example-Output-Malicious}
\end{figure}

Figures \ref{fig:CIC-Example-Output-Benign} and \ref{fig:CIC-Example-Output-Malicious} present the corresponding LLM responses for the attack scenario from Figure \ref{fig:CIC-Example-Prompt}, produced when prompted with the original attack description (pre-attack), and the adversarially perturbed description (post-attack), respectively. Both responses include an attack behavior analysis highlighting abnormal traffic features, along with mitigation suggestions tailored to the specified Raspberry Pi device. Initially, the framework benchmarked \textit{ChatGPT-o3} as the ideal LLM for this task, and in this work, we use its successor, \textit{ChatGPT-5 Thinking} \cite{openai2025-thinking}. This prompt is executed for each class in the combined set of 28 attack classes across both datasets, first with the original attack description and then with its adversarially perturbed version, and the resulting outputs are collected for comparative performance evaluation.

\subsubsection{Response Evaluation with Judge LLMs}
The \textit{evaluation prompt} assigns judge LLMs the role of cybersecurity professionals tasked with scoring two responses generated by \textit{ChatGPT-5 Thinking} for the same attack scenario. One response corresponds to when the original attack description was used in the scenario prompt, and the other corresponds to when the adversarially perturbed description was used. To avoid attacking the judge LLM and to ensure accurate evaluation, the attack scenario in the evaluation prompt is always provided with the original description.

\begin{figure}[!t]
    \centering
    \includegraphics[width=1\linewidth]{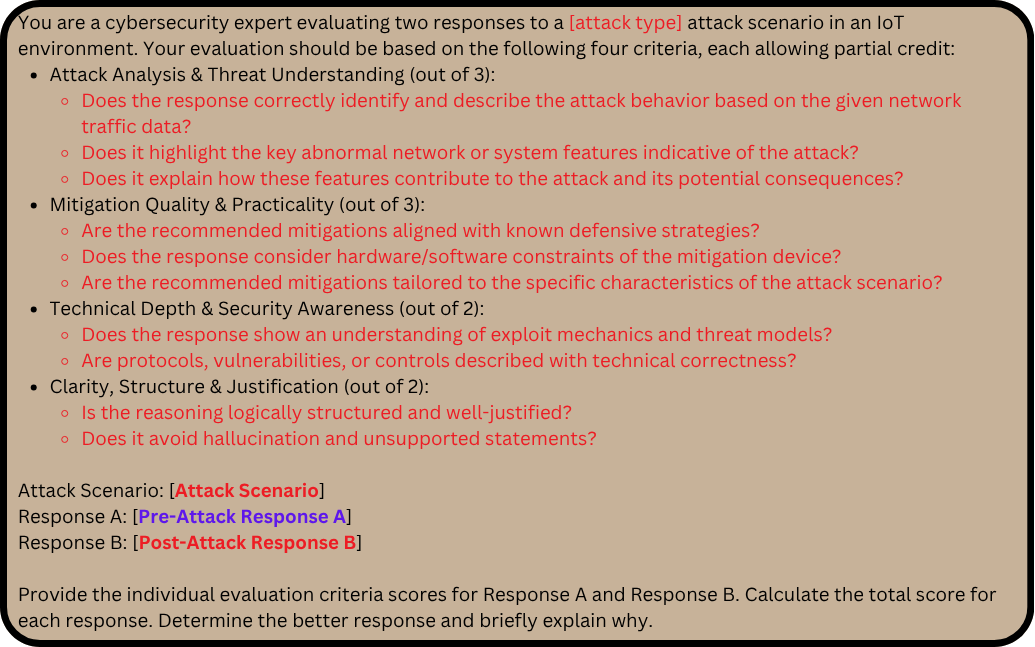}
    \caption{Evaluation Prompt Template}
    \label{fig:Evaluation-prompt}
\end{figure}

An example evaluation prompt is shown in Fig. \ref{fig:Evaluation-prompt}. The judge LLMs score the responses using predefined metrics covering attack analysis and threat understanding, mitigation quality and practicality, technical depth and security awareness, as well as clarity and organization. Scores across the metrics are aggregated into a total, and each judge provides a short justification for its scoring decision. To ensure a fair comparative analysis, and to achieve an accurate measurement of the adversarial attack impact in \textit{ChatGPT-5 Thinking}, we employ the same set of judge LLMs used in the framework, or their direct successors. The employed judge LLMs are ChatGPT-5 Instant, Mixtral 8x7B Instruct, Gemini 2.5 Flash, Meta Llama 4, TII Falcon H1 34B Instruct, DeepSeek-V3.2, xAI Grok 4, and Claude Sonnet 4.5.

\section{Performance Evaluation}
\label{sec:eval}
This section presents the metrics and results for fine-tuning the \textit{BERT} surrogate target model and for evaluating the adversarial attack on \textit{ChatGPT-5 Thinking} within the framework.

\subsection{Evaluation Metrics}
\label{sec:evaluation-metrics}
Our framework uses two sets of quantifying evaluation metrics. The first set of metrics measures the performance of the fine-tuned \textit{BERT} surrogate target model trained on the description dataset generated in section \ref{sec:AttackDescriptionDatasetGeneration}. We report \textit{Precision}, \textit{Recall}, and \textit{F1-score} of the fine-tuned \textit{BERT} model. The dataset spans 18 attack classes rather than the 28 classes found across both datasets, since several dataset classes share a common description. The second set of metrics measures the performance of \textit{ChatGPT-5 Thinking} before and after the adversarial attack. To ensure a fair measurement of the adversarial attack impact, we use the same scoring rubric to assess the pre-attack and post-attack responses, which consists of four metrics that total ten points across ten criteria.

The first metric, \textit{Attack Analysis and Threat Understanding}, is scored on a three-point scale and assesses correct identification of the attack, accurate description of abnormal features, and their impact. The second metric, \textit{Mitigation Quality and Practicality}, is scored on a three-point scale and assesses relevance and applicability of countermeasures with attention to device and deployment constraints. The third metric, \textit{Technical Depth and Security Awareness}, is scored on a two-point scale and assesses the model's understanding of mechanisms, protocols, vulnerabilities, and controls. The fourth and final metric, \textit{Clarity, Structure, and Justification}, is scored on a two-point scale and assesses organization, sound reasoning, and the absence of unsupported claims. These metrics are evaluated by us (authors) and the judge LLMs, and we report the scores before and after the adversarial attack.

\subsection{Results}
\subsubsection{Surrogate Target Model Classification Results}
Table \ref{tab:bert_finetune_results} presents the per-class text-based attack classification results for the \textit{BERT} model, which achieved an accuracy of 0.9722 and an F1-score of 0.9729. Most classes achieved a perfect F1-score, while lower performance was observed for closely related TCP-oriented floods. Specifically, TCP SYN Flood, TCP Flood, and SYN Flood showed reduced scores because they are closely related in behavior and wording. So, the fine-tuned \textit{BERT} model demonstrated near-perfect classification performance based on the attack descriptions.

\begin{table}[!t]
\caption{\textit{BERT} Classification Report}
\label{tab:bert_finetune_results}
\begin{center}
\begin{tabular}{|l|c|c|c|c|}
\hline
\textbf{Attack Class} & \textbf{Precision} & \textbf{Recall} & \textbf{F1-score} & \textbf{Support} \\
\hline
\rowcolor{rowgray}
Backdoor & 1.0000 & 1.0000 & 1.0000 & 6 \\ \hline
HTTP Flood & 1.0000 & 1.0000 & 1.0000 & 6 \\ \hline
\rowcolor{rowgray}
ICMP Flood & 1.0000 & 1.0000 & 1.0000 & 6 \\ \hline
SYN Flood & 1.0000 & 0.8333 & 0.9091 & 6 \\ \hline
\rowcolor{rowgray}
TCP Flood & 0.7143 & 0.8333 & 0.7692 & 6 \\ \hline
UDP Flood & 1.0000 & 1.0000 & 1.0000 & 6 \\ \hline
\rowcolor{rowgray}
DNS Spoofing & 1.0000 & 1.0000 & 1.0000 & 6 \\ \hline
Dictionary Brute Force & 1.0000 & 1.0000 & 1.0000 & 6 \\ \hline
\rowcolor{rowgray}
MITM & 1.0000 & 1.0000 & 1.0000 & 6 \\ \hline
ARP Spoofing & 1.0000 & 1.0000 & 1.0000 & 6 \\ \hline
\rowcolor{rowgray}
Password Cracking & 1.0000 & 1.0000 & 1.0000 & 6 \\ \hline
OS Scanning & 1.0000 & 1.0000 & 1.0000 & 6 \\ \hline
\rowcolor{rowgray}
Port Scanning & 1.0000 & 1.0000 & 1.0000 & 6 \\ \hline
SQL Injection & 1.0000 & 1.0000 & 1.0000 & 6 \\ \hline
\rowcolor{rowgray}
TCP SYN Flood & 0.8333 & 0.8333 & 0.8333 & 6 \\ \hline
Uploading & 1.0000 & 1.0000 & 1.0000 & 6 \\ \hline
\rowcolor{rowgray}
Vulnerability Scanning & 1.0000 & 1.0000 & 1.0000 & 6 \\ \hline
Cross-Site Scripting & 1.0000 & 1.0000 & 1.0000 & 6 \\ \hline
\rowcolor{rowgray}
\textbf{Macro Average} & 0.9749 & 0.9722 & 0.9729 & 108 \\ \hline
\textbf{Weighted Average} & 0.9749 & 0.9722 & 0.9729 & 108 \\ \hline
\rowcolor{rowgray}
\textbf{Accuracy (\%)} & \multicolumn{4}{c|}{\textbf{97.22}} \\
\hline
\end{tabular}
\end{center}
\end{table}

\subsubsection{Adversarial Attack Results}
To demonstrate the human expert evaluation of the adversarial attack on \textit{ChatGPT-5 Thinking}, we examine the pre-attack and post-attack responses for the Port Scanning attack scenario in Figure \ref{fig:CIC-Example-Prompt}, as shown in Figures \ref{fig:CIC-Example-Output-Benign} and \ref{fig:CIC-Example-Output-Malicious}, respectively. We illustrate this sample because its adversarial attack description required only five word substitutions from \textit{TextFooler}, the fewest among all samples, while maintaining a cosine similarity of 0.7631 to the original. In terms of attack analysis and threat understanding, the pre-attack response correctly identified and described the attack using key network-level indicators such as high packet emission rates, inbound-only traffic, elevated TCP flag volumes, as well as the absence of legitimate TCP handshakes. By contrast, the post-attack response surfaced the same indicators and referenced ports, but then shifted its attack behavior analysis towards exposed interfaces instead of explicitly connecting the traffic indicators to the port scanning attack, reflecting the context manipulation in the perturbed description. As for mitigation strategies, both responses proposed known defenses such as firewall controls and network Intrusion Prevention System (IPS) deployment \cite{mitre_t1046}, and they can be tailored to the Port Scanning scenario while considering the resource constraints of the target Raspberry Pi device. Unbiased by the perturbed attack description, the pre-attack response went further by including Port Scanning-specific defenses such as the Port Scan Attack Detector (PSAD), tailoring its output to the attack scenario at hand. Moreover, the post-attack response omitted full implementation code/configuration snippets for a few of the mitigations it proposed, such as Fail2Ban and network segmentation. Both responses demonstrated technical depth and security awareness, and were provided in a logical, hallucination-free, well-justified manner. Based on the evaluation criteria, \textit{ChatGPT-5 Thinking}'s pre-attack response scored a perfect 10 out of 10, while its post-attack response scored 8 out of 10.

\begin{figure}
    \centering
    \includegraphics[width=1\linewidth]{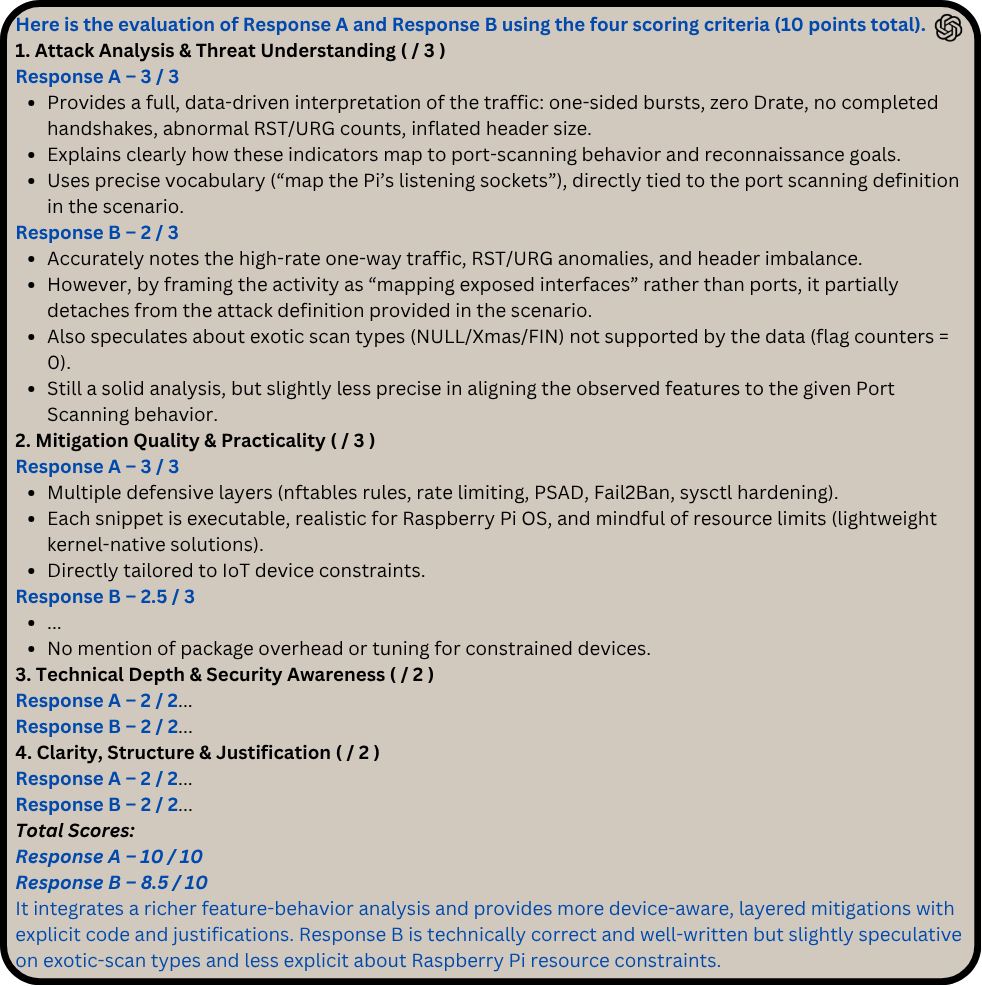}
    \caption{Example Judge LLM Response by ChatGPT-5 Instant}
    \label{fig:ChatGPT-5-Instant-Response}
\end{figure}

Figure \ref{fig:ChatGPT-5-Instant-Response} presents the evaluation response produced by \textit{ChatGPT-5 Instant}, demonstrating the judge LLM evaluation process by highlighting gaps in areas such as network traffic feature analysis, technicality, and mitigation suggestion. The evaluation reveals that Response A, which is \textit{ChatGPT-5 Thinking}'s pre-attack response, outperforms Response B, which is its post-attack response, receiving a perfect total score of 10 out of 10, compared to 8.5 out of 10. The judge LLM highlights reduced alignment in Response B, noting weaker linkage between observed features and the described behavior, fewer device-aware mitigations, and less explicit justification for recommendations. In contrast, Response A maintains precise feature behavior mapping and presents a larger suite of layered controls that include attack-specific countermeasures that are realistic for the Raspberry Pi 4 environment. The scores indicate that the attack degraded the \textit{Attack Analysis and Threat Understanding}, as well as the \textit{Mitigation Quality and Practicality} of \textit{ChatGPT-5 Thinking}.

\begin{figure}
    \centering
    \includegraphics[width=1\linewidth]{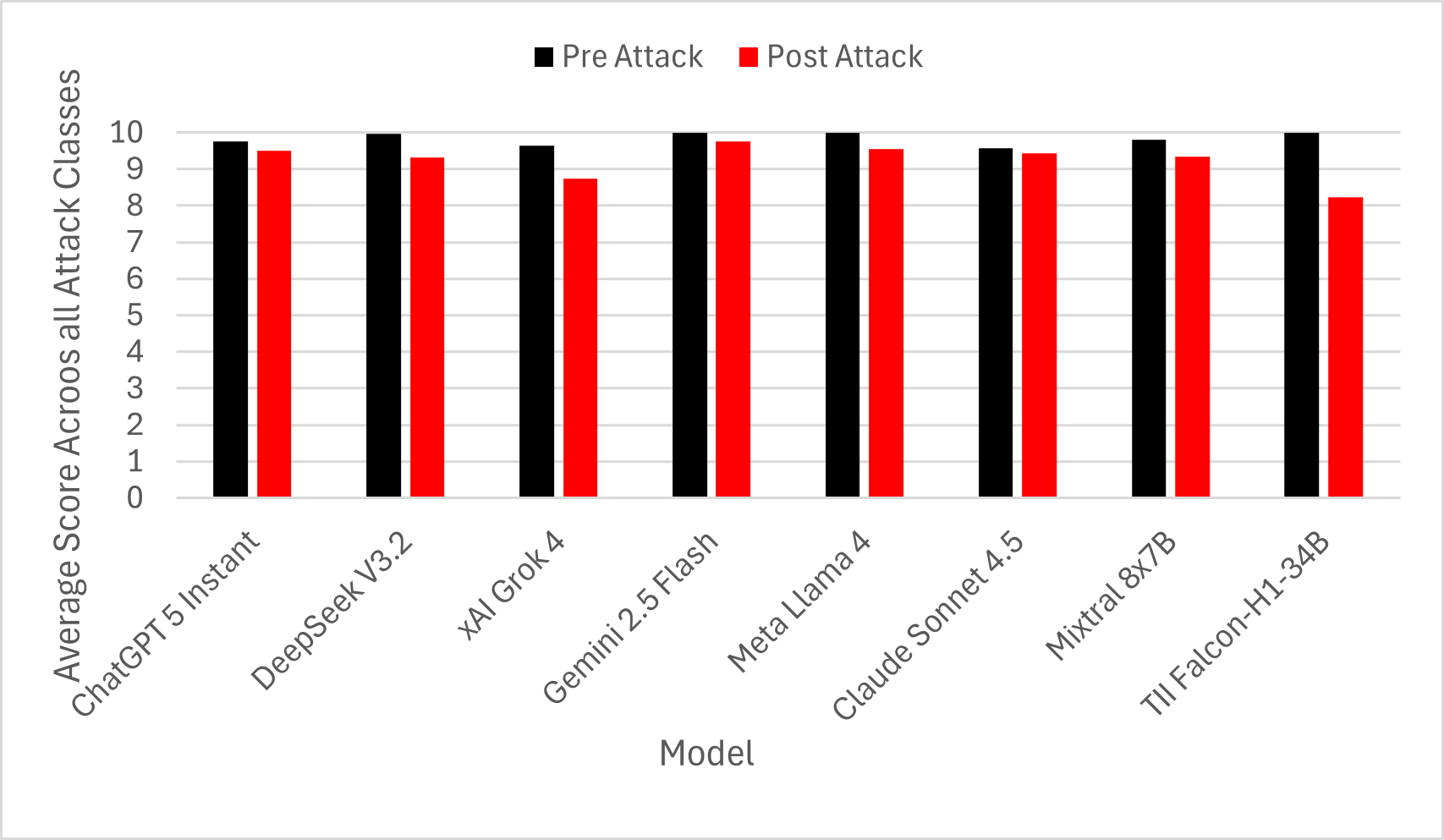}
    \caption{\textit{ChatGPT-5 Thinking} Performance (Edge-IIoTset Dataset)}
    \label{fig:edgeiiot_bar_chart}
\end{figure}

\begin{figure}
    \centering
    \includegraphics[width=1\linewidth]{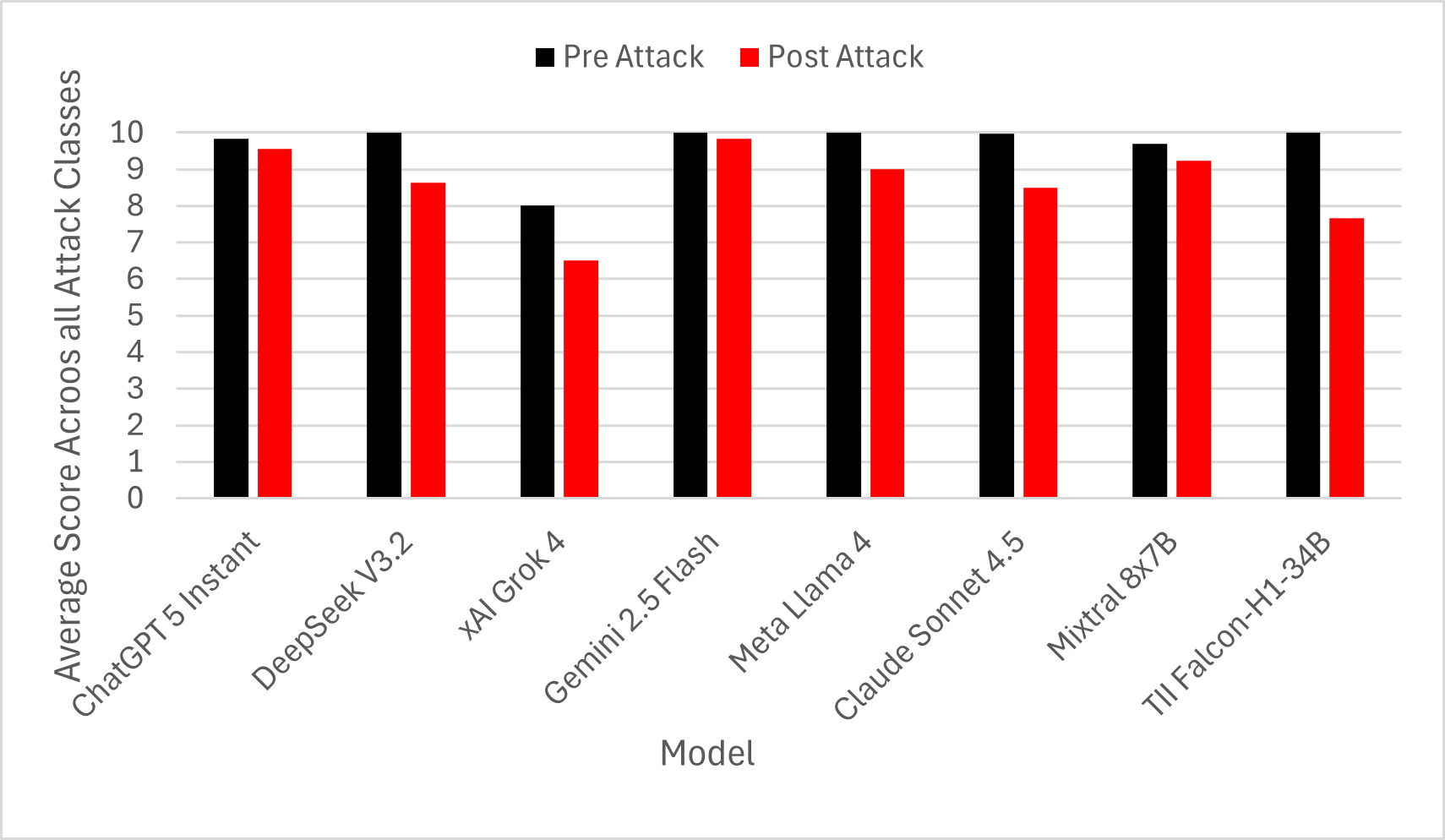}
    \caption{\textit{ChatGPT-5 Thinking} Performance (CICIoT2023 Dataset)}
    \label{fig:ciciot_bar_chart}
\end{figure}

We applied this evaluation to every attack scenario across both datasets using the four metrics in section \ref{sec:evaluation-metrics}, with scoring performed jointly by us (authors) and the judge LLMs. Figure \ref{fig:edgeiiot_bar_chart} and \ref{fig:ciciot_bar_chart} display average scores achieved by \textit{ChatGPT-5 Thinking} before and after the adversarial attack, across all attack classes for Edge-IIoTset and CICIoT2023 datasets, as assigned by each judge LLM. On the Edge-IIoTset, the pre-attack responses consistently scored higher, achieving perfect scores from Gemini 2.5 Flash, Meta Llama 4, and TII Falcon-H1-34B-Instruct. On the other hand, the post-attack responses demonstrated lower performance across all judge LLMs, receiving the lowest average score of 8.23 from TII Falcon-H1-34B-Instruct. On CICIoT2023, the pre-attack responses also achieved higher scores, achieving perfect scores from DeepSeek V3.2, Gemini 2.5 Flash, Meta Llama 4, and TII Falcon-H1-34B-Instruct. Yet again, the post-attack responses demonstrated lower performance across all judge LLMs, receiving the lowest average score from TII Falcon-H1-34B-Instruct.

\begin{figure}
    \centering
    \includegraphics[width=1\linewidth]{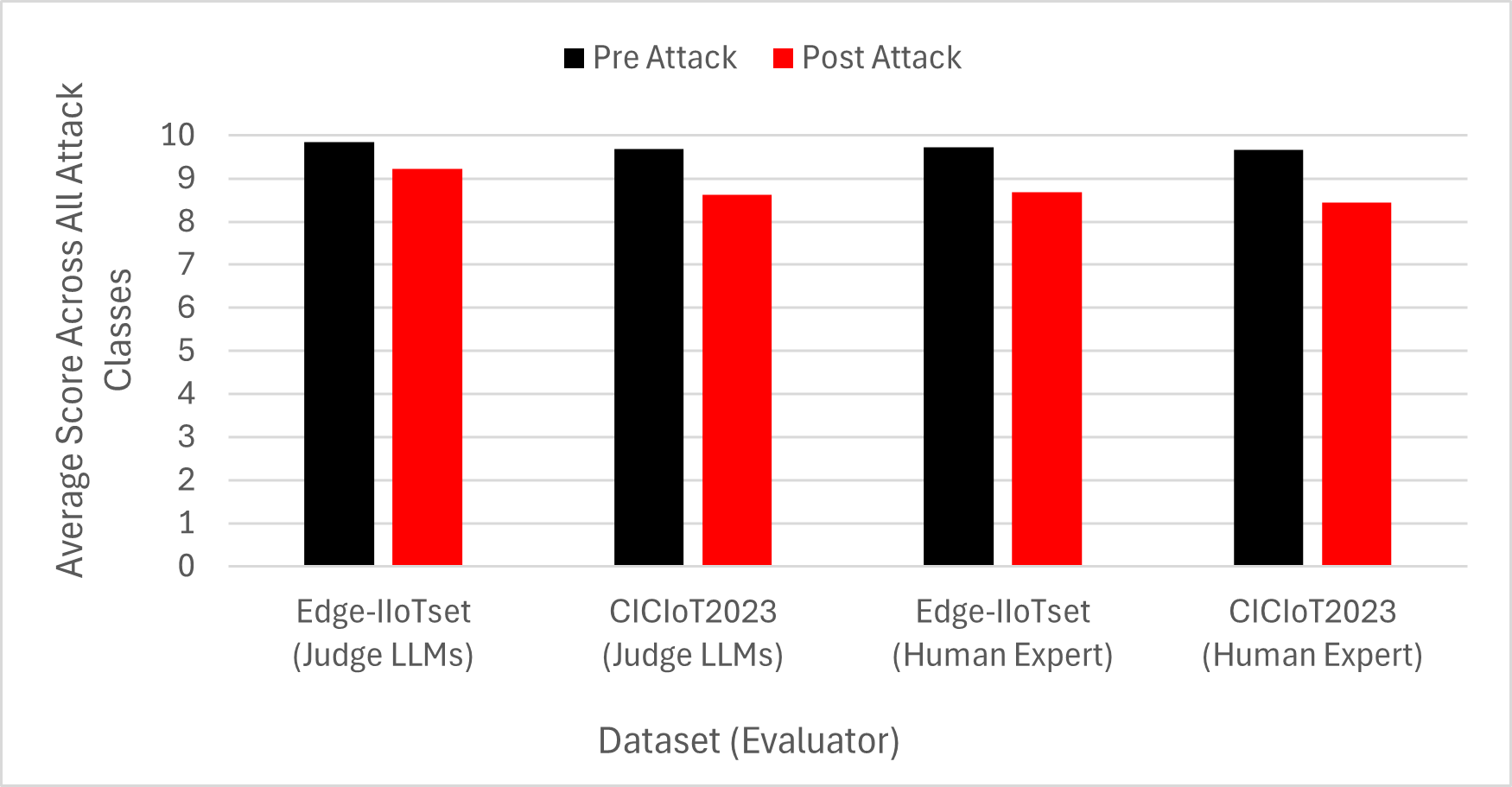}
    \caption{\textit{ChatGPT-5 Thinking} Overall Performance}
    \label{fig:overall-performance}
\end{figure}

Figure \ref{fig:overall-performance} compares average scores for the pre-attack and post-attack responses across all attack classes, measuring the effect of the adversarial attack based on aggregated results from the ensemble of judge LLMs and human expert assessments. On the Edge-IIoTset, the pre-attack responses achieved an average score of 9.85 from judge LLMs and 9.73 from human expert assessments, while the post-attack responses achieved 9.23 and 8.69, respectively. On the CICIoT2023, the pre-attack responses achieved an average score of 9.69 from judge LLMs and 9.67 from human expert assessments, while the post-attack responses achieved 8.62 and 8.43, respectively. Overall, the results reflect the success of the adversarial attack in degrading the performance of \textit{ChatGPT-5 Thinking}'s attack behavior analysis and mitigation suggestion, with higher effectiveness on the CICIoT2023 dataset environment. Despite \textit{ChatGPT-5 Thinking} being the latest and most capable model in its family, our RAG-targeted transfer-learning attack achieved measurable, albeit limited, degradation in model performance. These results provide clear evidence that our adversarial attack is still capable of degrading the performance even in state-of-the-art LLMs.

\section{Conclusion and Future Works}
\label{sec:summary}
In this work, we carry out an adversarial poisoning attack against an LLM-based IoT attack analysis and mitigation framework to test its adversarial robustness. We construct an attack description dataset and use it in a transfer-learning-based data-poisoning attack that minimally perturbs texts (attack descriptions) using \textit{TextFooler} against \textit{BERT}, a surrogate target model, to corrupt the attack description knowledge base in the RAG component of the framework. We then compare the pre-attack and post-attack performance of \textit{ChatGPT-5 Thinking} on samples derived from the Edge-IIoTset and CICIoT2023 datasets using quantitative performance measurement metrics. Our results show that introducing word-level perturbations in the attack descriptions consistently degrades the quality of LLM-generated attack behavior analyses and mitigation suggestions. Future work should explore combining our RAG data-poisoning attack with controlled perturbations to network-traffic features to further increase the misalignment between retrieved context and observed features, enabling a systematic study of coordinated multi-signal attacks and their impact on adversarial robustness.

\section*{Acknowledgment}
Authors would like to thank Dr. Elmahedi Mahalal for his inputs during the early stages of this study. This work is partially supported by the NSF grants 2416990 and 2346001, and NSA award H98230-24-1-0102 at Tennessee Tech University.

\bibliographystyle{ieeetr}
\bibliography{references}
\end{document}